\documentclass{pos}
\usepackage{amssymb,fontenc,times,mathptmx,graphicx}
\usepackage{amsmath}
\newcommand{\beq}{\begin{equation}}
\newcommand{\eeq}{\end{equation}}
\newcommand{\beqa}{\begin{eqnarray}}
\newcommand{\eeqa}{\end{eqnarray}}
\newcommand{\bseq}{\begin{subequations}}
\newcommand{\eseq}{\end{subequations}}

\def\r{{\bf r}}
\def\z{{\bf z}}

\def\q{{\bf q}}

\def\0{{\bf 0}}

\def\cal{\mathcal}
\def\lsim{\raise0.3ex\hbox{$<$\kern-0.75em\raise-1.1ex\hbox{$\sim$}}}
\def\gsim{\raise0.3ex\hbox{$>$\kern-0.75em\raise-1.1ex\hbox{$\sim$}}}
\title{Real and imaginary-time quarkonium correlators in a hot plasma}

\ShortTitle{In-medium quarkonium correlators }

\author{\speaker{Andrea Beraudo}\\
        ECT* (Trento) and Universit\'a  di Torino\\
        E-mail: \email{beraudo@to.infn.it}}

\author{Jean Paul Blaizot\\
        Saclay, SPhT\\
        E-mail: \email{Jean-Paul.Blaizot@cea.fr}}

\author{Claudia Ratti\\
        Stony Brook\\
        E-mail: \email{claudia.ratti@ph.tum.de}}
\abstract{The possibility of describing the behavior of a $Q\overline{Q}$ pair in a hot plasma in terms of an effective potential is investigated. It is shown that as long as medium effects can be embodied in a gaussian action, like in the QED case, the $Q\overline{Q}$ propagator obeys a closed temporal evolution equation whose large-time behavior is governed by an effective potential. The latter, beside screening, displays also an imaginary part related to collisions.}

\FullConference{8th Conference Quark Confinement and the Hadron Spectrum\\
                 September 1-6, 2008\\
                 Mainz. Germany}

\begin{document}

\section{Introduction}
More than 20 years ago Matsui and Satz \cite{Mat} proposed the $J/\psi$ suppression due to Debye screening as a signature of the onset of deconfinement, triggering a lot of experimental efforts to measure its yields in high-energy nucleus-nucleus collisions.
On the theory side in the last few years quite precise lattice data became available, both for the $Q\overline{Q}$ free-energies \cite{zan} and the in-medium correlators of quarkonia \cite{datta}. This, hopefully, should make possible a more solid quantitative study of the problem. Quite surprisingly the spectral functions extracted from the temporal correlators of quarkonia turned out to display well-defined peaks, reflecting the presence of bound/resonant states till temperatures of order $2T_c$, at least in the s-wave channels.
Various attempts were done in order to interpret both the above large melting temperatures \cite{wong,albe1} and the charmonia correlators themselves \cite{mocsy,albe2} in terms of screened potential models.
The latter represents a quite economic way of accounting for medium effects, which is often employed to describe the interaction between charged particles in a polarizable medium.
However, beside checking the numerical agreement of the findings obtained in the two approaches, it would be of interest to see such a medium-modified
$Q\overline{Q}$ potential arising from a first principle calculation. Different papers appeared quite recently addressing this issue \cite{lai1,berry,bra}.
Here we wish to answer a few very general questions concerning the description of a $Q\overline{Q}$ pair in a hot plasma: whether it is possible to give a solid theoretical basis to the concept of an effective in-medium potential; if so, which is its link with the $Q\overline{Q}$ free-energy obtained from the imaginary-time propagator of a static pair of heavy quarks; finally, whether, beside the screening of the interaction it is possible, within the same framework, to account also for other effects, like the collisional damping.
\section{The $Q\overline{Q}$ propagator in the complex time plane}
We start our investigation from the following propagator
\beq\label{eq:def}
G^>(t,\!\r_1;t,\!\r_2|0,\!\r_1';0,\!\r_2')\!\equiv\!
\langle\chi(t,\!\r_2)\psi(t,\!\r_1)
\psi^\dagger(0,\!\r_1')\chi^\dagger(0,\!\r_2')\rangle
\!\equiv\!\langle J(t,\r_1,\r_2)J^\dagger(0,\r_1',\r_2')\rangle
\eeq
of a $Q\overline{Q}$ pair created at time $0$ and annihilated at time $t$.
From its spectral decomposition it follows that $G^>(t)$ is an
analytic function of the (complex) time $t$ in the strip $-\beta\le\mbox{{\rm Im }}t\le0$.
In the case of static quarks the above propagator reduces to
\beq
G_{M=\infty}^>(t,\!\r_1;t,\!\r_2|0,\!\r_1';0,\!\r_2')\!\equiv\!
\delta(\r_1-\r_1')\delta(\r_2-\r_2')\overline{G}(t,\r_1-\r_2).
\eeq
The analyticity of $G^>(t)$ allows to follow its evolution for imaginary times. In particular, in the static case, its value at $\tau\!=\!\beta\!\equiv1/T$ 
\beq\label{freeenergy0}
\overline{G}(t\!=\!-i\beta,\r_1-\r_2)=e^{-\beta\Delta F_{Q\overline{Q}}(\r_1-\r_2)},
\eeq
gives the change of free-energy occurring once a $Q\overline{Q}$ pair is added into a finite-temperature medium. The latter is also the quantity provided by the lattice-QCD simulations of Polyakov line correlators\footnote{The evolution of a static quark from $\tau\!=\!0$ to $\tau\!=\!\beta$ is in fact described by a Polyakov line.} \cite{mclerr}, hence its interest in the present contest.

In \cite{lai1} the authors proposed the following strategy to properly define a \emph{real-time} potential. One assumes that, for static quarks, $G^>(t)$ obeys an equation like (with simplified notation)
\beq\label{eq:ans1}
[i\partial_t-V(t,r)]G_{M=\infty}^>(t)=0.
\eeq
One can then evaluate $G^>(t)$ at the lowest order of the HTL resummed perturbative expansion as  $G^>(t)=G_{(0)}^>(t)+G_{(2)}^>(t)+\dots$, plug it into Eq. (\ref{eq:ans1}), identifying in such a way the leading contribution to the effective real-time $Q\overline{Q}$ potential
\beq
V(t,r)=V^{(2)}(t,r)+\dots
\eeq
The latter is then employed in the dynamical problem with finite-mass quarks,
\beq\label{eq:ans2}
[i\partial_t-T-V^{(2)}(t,r)]G^>(t)=0,
\eeq
implicitly assuming that also in this case the pair propagator obeys a closed  Schr\"odinger equation.

The above procedure is \emph{far from trivial}. In general in fact the temporal evolution of a $n-particle$ propagator is part of an infinite hierarchy of coupled equations. Is it possible to find some physical systems which share many relevant features with the QGP, but for which one can get a closed Schr\"odinger equation for the heavy-pair propagator $G^{>}(t)$ from a first-principle calculation?
A hot QED plasma of photons, electrons and positrons will answer to our purposes and will be the subject of the next section. 
\section{The $Q\overline{Q}$ propagator in a hot QED plasma}
We concentrate on the case of a static pair and in order to evaluate its evolution in the complex-time plane we proceed as follows. The propagator in a given background configuration of the gauge field is given by the product of two Wilson lines
\beq
\overline{G}_A(t,\r_1,\r_2)\!=
\exp\left(ig\!\int_{0}^t dt'A_0(\r_1,t')\right)
\exp\left(-ig\!\int_{0}^{t} dt'A_0(\r_2,t')\right)
\!\equiv\exp\left(i\!\int d^4z\, J^\mu(z)A_\mu(z)\right).
\eeq
One should then average over all the possible field configurations with an action accounting for medium effects:
\beq
\overline{G}(t,\r_1-\r_2)
=Z^{-1}\int[{\cal D}A]\overline{G}_A(t,\r_1,\r_2)\,e^{iS[A]}.
\eeq
For the latter a convenient choice is the HTL effective action, which allows to properly include the most relevant medium corrections to the propagation of long wave-length modes ($\lambda_{\rm \,soft}\!\sim\!1/gT$). It can be expressed in terms of the time-ordered (along the usual Schwinger-Keldysh contour $C$ in the complex-time plane described in \cite{berry}) photon propagator
\beq
D_{\mu\nu}(x-y)\equiv i\,\theta_C(x^0-y^0)
\langle A_\mu(x)A_\nu(y)\rangle
+i\,\theta_C(y^0-x^0)\langle A_\nu(y)A_\mu(x)\rangle,
\eeq
taken in the HTL approximation, and reads
\beq
S_C^{HTL}[A]=
-\frac{1}{2}\int_C d^4x\int_C d^4y\,A^\mu(x)\left(D^{-1}\right)
_{\mu\nu}^{HTL}(x-y)A^\nu(y).
\eeq
Being the above action gaussian allows to perform the functional integral exactly getting
\beq
\overline{G}(t,\r_1-\r_2)=
\exp\left[\frac{i}{2}\int_{C} d^4x\int_{C}d^4y\,
J^\mu(x)D_{\mu\nu}^{HTL}(x-y)J^\nu(y)\right].
\eeq
\subsection{Real-time propagation}
A $Q\overline{Q}$ pair propagating along the real-time axis is described by the current
\beq
J^\mu(z)=\delta^{\mu 0}\theta(z^0)\theta(t-z^0)
[g\delta(\z-\r_1)-g\delta(\z-\r_2)],
\eeq
which leads to 
\beq
\overline{G}(t,\r_1\!-\!\r_2)
\!=\!\exp\!\left[2ig^2\!\int\frac{d\omega}{2\pi}\!\int\frac{d\q}{(2\pi)^3}
\frac{1-\cos(\omega t)}{\omega^2}
\!\left(1\!-\!e^{i\q\cdot(\r_1-\r_2)}\right)\!
D_{00}(\omega,\q)\!\right],
\eeq
where the FT of the time-ordered propagator (which we take in Coulomb gauge) can be expressed in terms of the HTL photon spectral function as follows \cite{berry}:
\beq\label{eq:QEDsp}
D_{00}(\omega,\q)=\frac{-1}{\q^2}+\int_{-\infty}^{+\infty}\frac{dq^0}{2\pi}
\frac{\rho_{L}(q^0,\q)}{q^0-(\omega+i\eta)}+i\rho_{L}(\omega,\q)N(\omega)\,.
\eeq
The above exact exponentiation, \emph{occurring in the case of a gaussian action}, allows an immediate identification of the real-time potential $V(t,r)$ defined implicitly in Eq. (\ref{eq:ans1}). It is of interest, in particular, to consider the large-time behavior of the above propagator, which results governed by the static limit of the HTL photon propagator. One has:
\beq
\lim_{t\to +\infty}[i\partial_t-V_{\rm eff}(\r_1-\r_2)
]\overline{G}(t,\r_1-\r_2)=0,
\eeq
with the in-medium effective potential we were looking for given by
{\setlength\arraycolsep{1pt}
\beqa
%\begin{multline}
V_{\rm eff}(\r_1-\r_2)&\equiv& g^2\int\frac{d\q}{(2\pi)^3}\left(1-e^{i\q\cdot
(\r_1-\r_2)}\right)\Big[\frac{1}{\q^2+m_D^2}
\!-i\frac{\pi m_D^2 T}
{|\q|(\q^2+m_D^2)^2}\Big]\nonumber\\
{}&=&-\frac{g^2}{4\pi}\left[m_D+\frac{e^{-m_Dr}}{r}\right]
-i\frac{g^2T}{4\pi}\phi(m_Dr).
%\end{multline}
\eeqa}
The above potential, arising here from a first-principle calculation, embodies medium effects both in its real part, with the screening of the interaction and the heavy-quark self-energy correction (its finite value requires subtracting the vacuum Coulomb self-interaction), and in its imaginary part related to the \emph{soft collisions} with the plasma particles suffered by the heavy quarks. Note that the self-energy correction is crucial to ensure that medium effects vanish for very small separations and that an effective potential of the kind
\beq
V_{\rm eff}(r)=-\alpha m_D-\frac{\alpha}{r}e^{-m_D r}
\eeq
is known in solid-state physics as Ecker-Weitzel potential \cite{bla}, often employed in the study of excitons in semiconductors.

In \cite{lai1} the propagator in Eq. (\ref{eq:def}) was given a gauge invariant definition by joining the heavy quark fields with two Wilson lines. However for the study of the $t\!\to\!\infty$ behavior this is of no consequence since only the diagrams with all the vertices attached to the long sides of the resulting Wilson loop give the leading contribution.
The situation is different in considering the imaginary-time evolution, which occurs only till $\tau\!=\!\beta$. The latter will be addressed in the next section.
\subsection{Imaginary-time propagation}
We now consider the $Q\overline{Q}$ propagation along the imaginary-time axis. It can be expressed in terms of the Matsubara (HTL resummed) photon propagator
\beq
\overline{G}(-i\tau,\r_1\!-\!\r_2)=\exp\!
\left[g^2\!\int_0^\tau \!d\tau'\!\int_0^\tau \!d\tau''
\!\int\frac{d\q}{(2\pi)^3}\!\left(1\!-\!e^{i\q\cdot(\r_1\!-\!\r_2)}\right)
\!\Delta_{L}(\tau'\!-\!\tau'',\q)\!\right],
\eeq
obtaining
\begin{multline}
\overline{G}(-i\tau,\r_1-\r_2)=\exp\left\{g^2\!\int\frac{d\q}{(2\pi)^3}
\left(1\!-\!e^{i\q\cdot(\r_1-\r_2)}\right)\left(\frac{-1}{\q^2}+\int\frac{dq^0}{2\pi}
\frac{\rho_L(q^0,\q)}{q^0}\right)\tau\right\}\times\\
\times\exp\left\{g^2\!\int\frac{d\q}{(2\pi)^3}
\left(1\!-\!e^{i\q\cdot(\r_1-\r_2)}\right)\int\frac{dq^0}{2\pi}
\frac{\rho_L(q^0,\q)}{(q^0)^2}(e^{-q^0\tau}-1)(1+N(q^0))\right\}\times\\
\times\exp\left\{g^2\!\int\frac{d\q}{(2\pi)^3}
\left(1\!-\!e^{i\q\cdot(\r_1-\r_2)}\right)\int\frac{dq^0}{2\pi}
\frac{\rho_L(q^0,\q)}{(q^0)^2}(e^{q^0\tau}-1)N(q^0)\right\}.\label{eq:2imagtau}
\end{multline}
Its value at $\tau\!=\!\beta$
\beq
\overline{G}(-i\beta,\r_1\!-\!\r_2)
=\exp\left\{-\beta g^2\!\int\frac{d\q}{(2\pi)^3}
\left(\!1\!-\!e^{i\q\cdot(\r_1\!-\!\r_2)}\!\right)\frac{1}{\q^2+m_D^2}\right\}
\eeq
allows (after subtracting the divergent vacuum contribution to the heavy-quark self-energy) to identify the $Q\overline{Q}$ free-energy, getting:
\beq
\Delta F_{Q\overline{Q}}(r,T)=
-\frac{g^2m_D}{4\pi}-\frac{g^2}{4\pi}\frac{e^{-m_Dr}}{r}.
\eeq
Note that the correlation function from which the latter was obtained, and which represents a simplified version of the usual Polyakov-line correlator evaluated in lattice-QCD, is indeed a gauge-dependent quantity. This issue was carefully examined in \cite{phil1,phil2}, where the usual strategy of extracting an in-medium effective potential from the lattice color-singlet free-energy was criticized.
What we can say after our investigation is that, at least in the case of a QED plasma for which an exact exponentiation holds, \emph{the $Q\overline{Q}$ free-energy evaluated in the Coulomb gauge coincides with the real part of the effective potential governing the large-time evolution of the pair propagator}. On the other hand, by looking only at the imaginary-time correlator at $\tau\!=\!\beta$, one looses any information about the collisional broadening suffered by the pair.
\section{Conclusions}
 We addressed some very general aspects related to the propagation of a heavy $Q\overline{Q}$ pair (used as an external probe to study medium properties) in a hot plasma. For the sake of simplicity we focused on the QED case. Medium corrections to the propagation of long wave-length modes are correctly described by the HTL effective action, which, for a QED plasma, turns out to be gaussian. This allows to get, for a static heavy-quark pair, an exact exponentiation of the resummed photon propagator leading to a closed Schr\"odinger equation governing the temporal evolution of the $Q\overline{Q}$ propagator, with an effective potential containing both a real (related to screening) and an imaginary part (related to soft collisions, i.e. Landau damping).

At least two questions require further investigation. The first one is to study whether an exponentiation similar to the one we found holds also in the QCD case \cite{expo}. The second one is to consider the propagation of a \emph{finite-mass} $Q\overline{Q}$ pair, checking whether the ansatz of a closed Schr\"odinger equation for $G^>(t)$ is justified or a more refined strategy is required. This is presently the object of our efforts.  

\end{document}